\documentclass[
prd
,twocolumn%
,nofootinbib
,amssymb,superscriptaddress,aps
]{revtex4-2}
\usepackage{graphicx}
\usepackage{color}
\input{epsf}

\usepackage{natbib}

\usepackage{amsmath,amssymb}
\usepackage{bm}
\usepackage{times}
\usepackage{ulem}

\usepackage{hyperref}
\hypersetup{
    colorlinks=true,
    linkcolor=blue,
    filecolor=magenta,      
    urlcolor=cyan,
    citecolor=blue
}
\newcommand{\dalm}{\kern1pt\vbox{\hrule height 0.9pt\hbox{\vrule width 0.9pt
\hskip 2.5pt\vbox{\vskip 5.5pt}\hskip 3pt\vrule width 0.3pt}\hrule height 0.3pt}
\kern1pt}

\defcitealias{SMT24}{SMT24}
\begin{document}



\title{Effects of multidimensional treatment of gravity in simulations on supernova gravitational waves}

\author{Hajime Sotani}
\email{sotani@yukawa.kyoto-u.ac.jp}
\affiliation{Department of Mathematics and Physics, Kochi University, Kochi, 780-8520, Japan}
\affiliation{RIKEN Center for Interdisciplinary Theoretical and Mathematical Sciences (iTHEMS), RIKEN, Wako 351-0198, Japan}
\affiliation{Theoretical Astrophysics, IAAT, University of T\"{u}bingen, 72076 T\"{u}bingen, Germany}

\author{Bernhard M\"{u}ller}
\affiliation{Monash Centre for Astrophysics, School of Physics and Astronomy, 11 College Walk, Monash University, Clayton VIC 3800, Australia}

\author{Tomoya Takiwaki}
\affiliation{Division of Science, National Astronomical Observatory of Japan, 2-21-1 Osawa, Mitaka, Tokyo 181-8588, Japan}
\affiliation{Center for Computational Astrophysics, National Astronomical Observatory of Japan, 2-21-1 Osawa, Mitaka, Tokyo 181-8588, Japan}

\date{\today}

\begin{abstract}
Supernova explosions are expected as one of the promising candidates for gravitational wave sources. In this study, we examine the supernova gravitational waves, focusing on the multidimensional treatment of gravity in the simulation. For this purpose, we newly performed two-dimensional relativistic simulations with a nonmonopole (two-dimensional) potential and compared the resultant gravitational wave signals in the simulations with the frequencies of the proto-neutron stars with and without the Cowling approximation. Then, we find that the proto-neutron star frequencies with the Cowling approximation overestimate the gravitational wave frequencies.
On the other hand, the frequencies of the proto-neutron star oscillations with metric perturbations agree well with the gravitational wave signals in the simulations. Employing the new data, we derive a new fitting formula for the supernova gravitational wave frequencies with the two-dimensional gravitational potential, independently of the progenitor mass. Combining this new formula with the previous one derived from the Cowling approximation, we also derive the formula to predict the gravitational wave frequencies with a two-dimensional potential, using those with a monopole potential. 
\end{abstract}

%
\maketitle


\section{Introduction}
\label{sec:I}

The direct detection of gravitational waves from compact binary mergers opens a new avenue for extracting astrophysical information via gravitational waves, together with electromagnetic waves and neutrinos. In practice, observations of gravitational waves permit the determination of the masses of black holes and neutron stars, e.g.,~\cite{GW1,GW6}, and even the radius of neutron stars can be constrained through the tidal deformability constraint \cite{Annala18}. Since, owing to the development of detectors, the detection rate for the gravitational waves from binary mergers is increasing, one can anticipate more stringent constraints on neutron star properties form gravitational waves in the coming years. Another promising gravitational wave source next to the compact binary mergers may be the supernova gravitational waves \cite{Abdikamalov2022,Szczepanczyk2022,Arimoto2023,Mezzacappa2024}. Because the gravitational waves from core-collapse supernovae are much weaker than those from compact binary mergers, one can only detect the gravitational waves from supernovae that occur in our galaxy with current instruments \cite{Hayama2015,Szczepanczyk2023,Szczepanczyk2024,Bruel23}. However, the detection of gravitational waves from a Galactic supernova would mark the first event where electromagnetic waves \emph{and} neutrinos are simultaneously detected. 

Pending a galactic event, gravitational waves from supernovae currently need to be studied via numerical simulations (e.g., non-rotating models~\cite{Murphy09,MJM2013,Yakunin15,KKT2016,Andresen16,RMBVN19,VBR2019,Powell2019,Andersen21,andresen21,Kuroda2022,Vartanyan23,Jakobus23,MoriK2025}, models with black holes~\cite{Kuroda2018,Pan2018,Rahman2022}, magnetized and rotational models~\cite{CDAF2013,Takiwaki2017,Richers2017,andresen2019,PM2020,Shibagaki2020,Takiwaki2021,Matsumoto2022,Shibagaki2024,Powell2024a}, models including exotic particles~\cite{MoriK2023,MoriK2024}, memory effect \cite{vartanyan20,Choi2024}). These investigations showed that, as a primary signal, a ``ramp-up'' gravitational wave signal emerges after core bounce, whose frequencies increase with time from a few hundred Hz up to the kHz range within $\sim 1$ second. The frequency of the ramp-up signal was originally considered to be determined by the Brunt-V\"{a}is\"{a}l\"{a} frequency associated with the thermal and composition gradient at the proto-neutron star surface (or surface gravity ($g$-) modes) \cite{MJM2013,CDAF2013}. Subsequent analysis has illuminated the precise nature of the ramp-up signal and its frequency dependence and shown that the signal comes from the fundamental ($f$-) mode oscillations (or the $g$-mode oscillations with the different classification from that as usual) of the proto-neutron stars produced via core-collapse supernova (e.g.,~\cite{Bruel23,MRBV2018,SKTK2019,ST2020b,TCPF2018,TCPOF2019a,TCPOF2019b,STT2021,Bizouard21,mori23,wolfe23,rodriguez23,SMT24}). In addition to the ramp-up signal, gravitational wave signals associated with the standing accretion-shock instability outside the proto-neutron star, whose frequencies around 100 Hz (e.g.,~\cite{KKT2016,Andresen16,andresen2019,vartanyan20,andresen21,SMT24,Kawahara18,Takeda21,Lin23}), and the signature of the $g_1$-mode oscillations, whose frequencies are decreasing with time~\cite{Jakobus23,MRBV2018,Kawahara18}, have been found in simulations, although these additional signals seem to strongly depend on the supernova models. 

To reveal the physics behind the signal components associated with the proto-neutron star oscillations, linear analysis of the proto-neutron stars is quite useful. This technique is known as (gravitational wave) asteroseisomology~\cite{KS1999}, which is a similar method to seismology on Earth and helioseismology on the Sun. Since the oscillation signals from objects strongly depend on their interior properties and each specific oscillation mode can be excited due to the corresponding physics, one could inversely extract the corresponding physics by identifying the observed frequencies with specific modes~\cite{AK1996,AK1998}. For example, by identifying the frequencies of quasi-periodic oscillations observed in magnetars with the neutron star crustal oscillations, the crust equation of state (EOS) and/or neutron star mass and radius have successfully been constrained (e.g.,~\cite{GNHL2011,SNIO2012,SIO2016,SKS23,Sotani24}). Similarly, it has been suggested that the EOS for neutron star matter, neutron star mass, radius, and rotational properties could be constrained, once the gravitational waves from neutron stars are observed (e.g.,~\cite{STM2001,SH2003,SYMT2011,PA2012,DGKK2013,Sotani2020,SK21}). Furthermore, this technique is being adopted even for the supernova gravitational waves in preparation for a galactic event (e.g.,~\cite{MRBV2018,SKTK2019,ST2020b,TCPF2018,TCPOF2019a,TCPOF2019b,STT2021,FMP2003,ST2016,Camelio17,SKTK2017,SS2019,WS2019,Mezzacappa20,ST2020a,ST2020c}).

For examining the supernova gravitational waves using the technique of asteroseismology, a quasi-hydrostatic proto-neutron star model is required as a background model. Unlike cold neutron stars, the construction of proto-neutron star models is more complicated, because the stellar properties are determined with the finite temperature EOS, i.e., one has to consistently treat the distribution of entropy and electron fraction as well as that of pressure and density inside the star. These properties can be determined from dynamical numerical simulations of the core collapse and the subsequent explosion; knowledge of the nuclear EOS alone is insufficient. From such multi-dimensional numerical simulations, one can then construct spherically symmetric proto-neutron star models amenable to a perturbative analysis by averaging the properties in the angular direction(s). 

Another difficulty peculiar to supernova gravitational wave asteroseismology is the choice of the outer boundary conditions in the linear analysis, because dense matter still surrounds the proto-neutron stars, unlike cold neutron stars. So far, two different approaches for the outer boundary have been adopted by several groups. One is to employ same boundary condition as usual asteroseismology at the proto-neutron star surface defined by a specific surface density ($\sim 10^{11}$\,g/cm$^3$), where the Lagrangian perturbation of pressure is set to zero. With this approach, one may be able to classify the resultant frequencies into specific oscillation modes, although the frequencies may depend on the selection of the surface density. Nevertheless, at least the $f$- and $g_1$-modes, which are focused on in this study, are almost independent of how to select the surface density~\cite{MRBV2018,ST2020b}. In this study, we simply adopt this approach, assuming that the surface density is $10^{11}$\,g/cm$^3$. Another approach treats the whole region inside the shock radius by adopting the boundary condition that the Lagrangian displacement in the radial direction is zero at the shock radius~\cite{TCPF2018,TCPOF2019a,TCPOF2019b}. Considering that matter exists even outside the proto-neutron star, this approach may be plausible, while one has to newly reclassify the resultant frequencies into specific modes because the problem to solve is mathematically different from the standard asteroseismology problem due to the different boundary conditions. 
Moreover, rigorously including the region between the proto-neutron star and the shock would require the incorporation of advection and wave reflection and coupling at the shock.
Anyway, the ramp-up signals emerging in the numerical simulations correspond well to the $f$-mode determined with the first approach or the ${}^2g_2$-mode using the classification in the second approach.

A persistent issue has been a systematic discrepancy between the signals in the simulations and oscillation frequencies of the proto-neutron stars \cite{ST2020b,STT2021,TCPF2018,TCPOF2019a,TCPOF2019b,SMT24,Zha2024}. 
The origin of this deviation likely arises from the inconsistency of the treatment of gravity in numerical simulations and linear analysis. The numerical simulations have been done either with the Newtonian or with the general relativistic framework, while we did the linear analysis in the relativistic framework. The Newtonian framework with the approximate gravitational potential, which mimics the potential produced from the Tolman-Oppenheimer-Volkoff equations, is known as the effective general relativistic (effective GR) simulations \cite{Marek06}, and the results obtained in the effective GR are expected to be better than those with the simple Newtonian simulations because the relativistic effects are partially taken into account in the effective GR.
Nevertheless, as discussed in the Appendix of Ref.~\cite{ST2020b}, consistently reconciling the mismatch between the effective GR simulation and the relativistic linear analysis is still challenging. Indeed, it has long been known that effective potentials lead to different mode eigenfrequencies than GR even when they reproduce the GR hydrostatic very accurately \citep{Mueller08}. Therefore, it is crucial to employ comparable gravitational treatments in both simulations and linear analysis to avoid such discrepancies. Indeed, when fully relativistic simulations are used, the gravitational wave signals show good agreement with the PNS oscillation frequencies derived from relativistic linear analysis, as demonstrated by Sotani, M\"{u}ller, and Takiwaki (2024; hereafter SMT24) \cite{SMT24}. This consistency has also been confirmed in an effective GR case by Zha {\it et al.} (2024)~\cite{Zha2024}.

Another important issue in supernova gravitational wave asteroseismology concerns \emph{universal relations} for mode frequencies. Since the gravitational wave signals strongly depend on the supernova parameters, such as the progenitor mass and EOS for dense matter, it may be difficult to extract the physical information from the direct observation of the gravitational waves. So far, two universal relations for the supernova gravitational wave frequencies are proposed as a function of the proto-neutron star average density ($M_{\rm PNS}/R_{\rm PNS}^3$) \cite{STT2021} or the surface gravity of proto-neutron star ($M_{\rm PNS}/R_{\rm PNS}^2$) \cite{TCPOF2019b}, using the proto-neutron star mass, $M_{\rm PNS}$, and radius, $R_{\rm PNS}$. The relation with $M_{\rm PNS}/R_{\rm PNS}^3$ seems to be more suitable to discuss the supernova gravitational waves because this relation is independent of treatment of gravity, i.e., whether the effective GR or relativistic framework and also independent of the numerical interpolations~\citepalias{SMT24}.

However, these analyses have been done with numerical simulations with the approximation of monopole gravitational potential and the linear analysis with the Cowling approximation, where the metric is fixed during the stellar oscillations. That is, it is not clear how the supernova gravitational waves are well described by the universal relation(s) in the case of numerical simulations with a more realistic multi-dimensional treatment of gravity and linear proto-neutron star oscillation frequencies determined with the metric perturbations. In this study, we will investigate how well the gravitational wave signals in a new relativistic numerical simulation with two-dimensional gravity are tracked by the frequencies of proto-neutron star oscillations determined with metric perturbations (i.e., without the Cowling approximation). Then, we will also discuss the universal relation of the supernova gravitational waves.

This paper is organized as follows. In Sec.~\ref{sec:PNSmodel}, we briefly describe the supernova simulations newly performed in this study and the resultant PNS models as background models to examine a linear analysis.
In Sec.~\ref{sec:GW}, we calculate the oscillation modes of gravitational waves from the PNS with and without the Cowling approximation.
The models employed in this paper are summarized in Table~\ref{tab:ourmodels}.
Then, we compare them with the gravitational wave signals appearing in the numerical simulations and discuss the universal relations for the gravitational wave frequencies. We present our conclusions in Sec.~\ref{sec:Conclusion}. 

\begin{table}[htbp]
    \centering
    \caption{Summary of the models. See Sec.~\ref{sec:PNSmodel} and \ref{sec:GW} for the meaning of GRm, GR2D, Cowling, and metric perturbations.
    }
    \begin{tabular}{cccc}
\hline
\hline
Progenitor           & \multicolumn{2}{c}{Treatment of gravity} & Related figure \\
\hline
 & Background & Seismology  &\\
\hline
S12 & GRm  &Cowling & Fig.~8 in \citetalias{SMT24} \\
S15 & GR2D & Cowling & Fig.~\ref{fig:GWSp_20SFHO_cow}\\
S15 & GR2D & metric perturbations &  Fig.~\ref{fig:GWSp_20SFHO}\\
S20 & GRm  & Cowling & Fig.~8 in \citetalias{SMT24} \\
S20 & GR2D & Cowling &  Fig.~\ref{fig:GWSp_20SFHO_cow} \\
S20 & GR2D & metric perturbations & Fig.~\ref{fig:GWSp_20SFHO} \\
\hline
\hline
    \end{tabular}
    \label{tab:ourmodels}
\end{table}

\section{Background models}
\label{sec:PNSmodel}

In this study, to investigate the importance of the multi-dimensional effect in the gravitational potential (or metric), we perform new two-dimensional relativistic numerical simulations with the non-monopole (two-dimensional) gravitational potential for the $15M_\odot$ and $20M_\odot$ progenitor models (as used for models S15 and S20 in Ref.~\cite{WH2007}). 
The simulations have been performed using the \textsc{CoCoNuT-FMT}
supernova code \citep{Mueller2015,Mueller2019}, which couples general relativistic hydrodynamics with the stationary three-flavor fast-multigroup transport (FMT) scheme. The FMT scheme combines a two-stream Boltzmann closure at high optical depth and an algebraic closure at low optical depth for the zeroth neutrino moment, and includes general relativistic redshift and simplified corrections for special relativistic Doppler shift.  Neutrino reactions in the code include absorption, emission, and scattering by nuclei and nucleons (including the effect of high-density correlation \citep{Horowitz2017}) and bremsstrahlung for heavy-flavor neutrinos. Energy transfer in neutrino-nucleon scattering is included in an approximative fashion.
We adopt the SFHo EOS~\cite{SFHo}. The spacetime metric in these simulations is calculated using the extended conformal flatness condition (xCFC)~\cite{cordero2009}.

\begin{figure}[tbp]
\begin{center}
\includegraphics[scale=0.6]{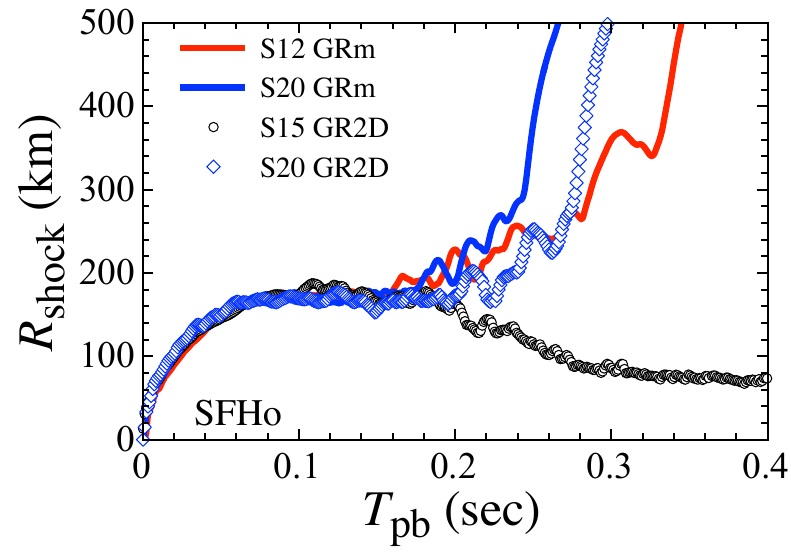}
\end{center}
\caption{
The evolution of the shock radius obtained from the relativistic simulations with S15 and S20, adopting the two-dimensional gravitational potential (S15~GR2D and S20~GR2D) is shown with circles and diamonds, respectively. The explosion time with S15 is around 0.8\,seconds after core bounce. For reference, the results of the relativistic simulations with S12 and S20, assuming the monopole approximation in gravitational potential (S12~GRm and S20~GRm) are also shown with the solid lines. 
}
\label{fig:Rshock}
\end{figure}

Figure~\ref{fig:Rshock} shows the evolution of the shock radius in the simulations using progenitors S15 and S20 (shown with circles and diamonds respectively), adopting the two-dimensional gravitational potential (S15~GR2D and S20~GR2D). 
For comparison, we also display the results obtained from the simulations with progenitors S12 and S20 with the solid lines, assuming the monopole potential approximation (S12~GRm and S20~GRm). 
The same abbreviation is used in Table~\ref{tab:ourmodels}.
Comparing the result of S20 GR2D with that of S20 GRm, due to the multipole effect in the gravitational potential, the explosion time was delayed a little ($\sim 40$\,msec). 
This delay is consistent with previous findings that small variations in microphysics or numerical methods can significantly impact explosion dynamics~\cite{just2018}.
In the case of S15 GR2D, the explosion is further delayed, occurring around 0.8 seconds after core bounce. This behavior is also consistent with previous results, such as Ref.~\cite{summa2016}, where shock revival times for S15 and S20 were reported to be 550\,msec and 292\,msec, respectively. That is, the shock revival is significantly delayed in S15, compared to S20.

\section{Gravitational wave asteroseismology}
\label{sec:GW}

To understand the gravitational wave signals appearing in the numerical simulations with the non-monopole (two-dimensional) gravitational potential, we perform a linear perturbation analysis with the Cowling approximation, where the fluid oscillations are considered with the fixed background metric, and also without the Cowling approximation, i.e., including the metric perturbations. By taking into account the metric perturbations, the eigenfrequencies become complex frequencies, where the real and imaginary parts correspond to the oscillation frequency and damping rate of the gravitational waves. However, the damping rate of the gravitational waves induced by the fluid oscillations is generally much smaller than the oscillation frequency, we simply neglect the damping rate (or the imaginary part of the complex frequency) and just consider the real frequency. Even with such an assumption, the frequencies are well determined at least on cold neutron stars, as in Ref.~\cite{Sotani22}. The perturbation equations and boundary conditions are the same as shown in Ref.~\cite{SKTK2019} with the Cowling approximation and in Ref.~\cite{ST2020c} without the Cowling approximation. 

\begin{figure*}[tbp]
\begin{center}
\includegraphics[scale=0.6]{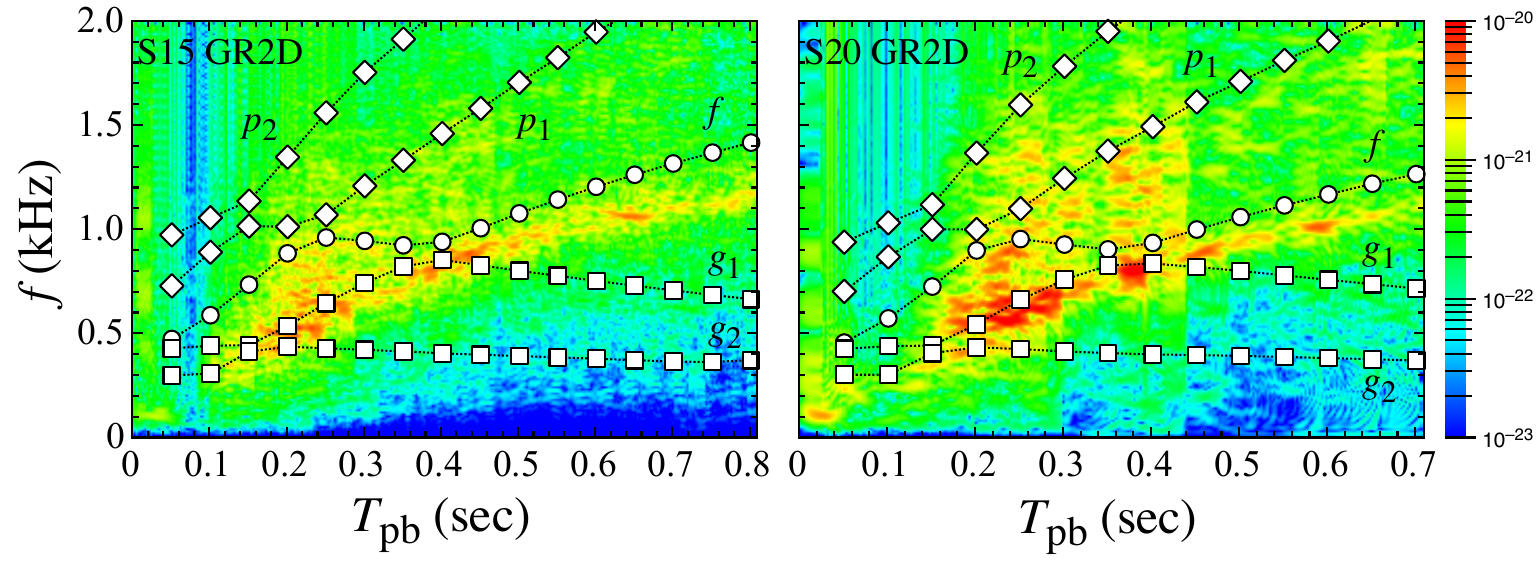}
\end{center}
\caption{
Comparison between the gravitational wave signals in the numerical simulation (background contour) and the proto-neutron star oscillation frequencies derived with the Cowling approximation (open marks) for the models with the $15M_\odot$ (left panel) and $20M_\odot$ progenitor (right panel) and SFHo EOS. The circles, squares, and diamonds correspond to the $f$-, $g_i$-, and $p_i$-modes with $i=1$ and 2.
The data is available in \url{https://doi.org/10.5281/zenodo.17168832}.
}
\label{fig:GWSp_20SFHO_cow}
\end{figure*}

First, as in the previous study, we consider the proto-neutron star oscillations with the Cowling approximation. Fig.~\ref{fig:GWSp_20SFHO_cow} compares the gravitational wave signals in the simulation (background contour) to the proto-neutron star oscillations (open marks), where the left and right panels correspond to the results with $15M_\odot$ and $20M_\odot$ progenitor models. From this figure, we find that the frequencies of the proto-neutron stars with the Cowling approximation are systematically larger than the gravitational wave signals in the simulations. Considering that the gravitational wave signals appearing in the simulations with monopole approximation agree well with the proto-neutron star oscillation frequencies determined with the Cowling approximation~\cite{SMT24}, the effect of non-monopole gravitational potential breaks this agreement.

\begin{figure}[tbp]
\begin{center}
\includegraphics[scale=0.6]{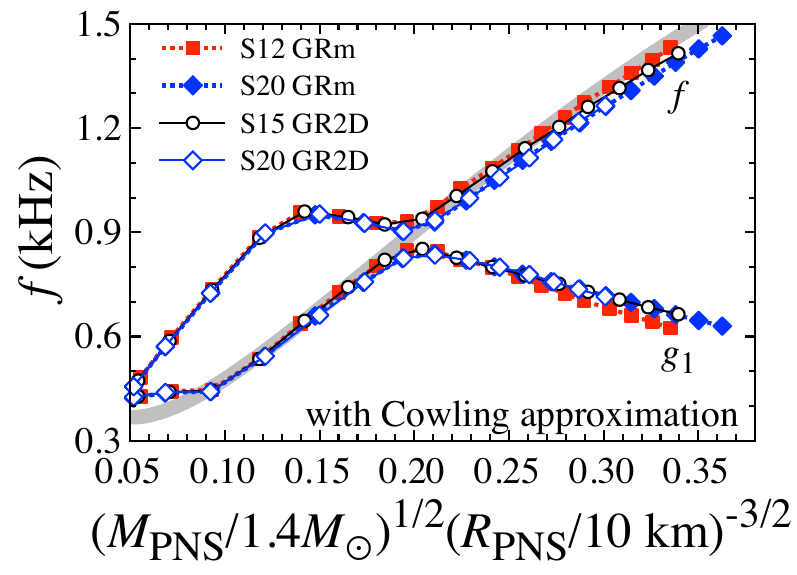}
\end{center}
\caption{
The $f$-  and $g_1$-mode frequencies determined via the eigenvalue problem with the Cowling approximation are compared with the empirical relation (thick solid line) given by Eq.~(\ref{eq:STT21}), which is originally derived in~\cite{STT2021}. The open marks denote the results obtained in this study with $15M_\odot$ and $20M_\odot$ progenitor models, while the frequencies obtained in the previous studies~\citetalias{SMT24} with $12M_\odot$ and $20M_\odot$ progenitor models, adopting the monopole approximation in gravitational potential, are also shown with the filled marks for reference.
}
\label{fig:fx-uni}
\end{figure}

Nevertheless, we find that the proto-neutron star oscillation frequencies determined with the Cowling approximation on the background models constructed with the simulation with non-monopole potential are well expressed using the universal relations, derived in \cite{STT2021} as
\begin{equation}
   f_{\rm Cow} {\rm (kHz)} = -1.410 -0.443 \ln(x) + 9.337x -6.714x^2, \label{eq:STT21}
\end{equation}
where $x$ is the normalized proto-neutron star average density defined with the proto-neutron star mass, $M_{\rm PNS}$, and radius, $R_{\rm PNS}$, by 
\begin{equation}
   x \equiv \left(\frac{M_{\rm PNS}}{1.4\,M_\odot}\right)^{1/2}\left(\frac{R_{\rm PNS}}{10\,{\rm km}}\right)^{-3/2}. \label{eq:xx}
\end{equation}
We focus only on the universal relation as a function of the stellar average density in this study, while the relation as a function of surface gravity defined as $M_{\rm PNS}/R_{\rm PNS}^2$ has also been suggested \cite{TCPOF2019b}. This is because the relation as a function of the stellar average density seems to be more relevant to discuss the supernova gravitational waves, where the relation is almost independent of the treatment of gravity and numerical interpolations~\cite{SMT24}.
In Fig.~\ref{fig:fx-uni}, we show the frequencies expected with this universal relation with the thick solid line and the proto-neutron star frequencies of the $f$- and $g_1$-modes determined with the Cowling approximation with the open marks with solid lines, i.e., circles for S15~GR2D and diamonds for S20~GR2D, as a function of the square root of the proto-neutron star average density. For reference, we also show the proto-neutron star frequencies determined with the Cowling approximation, adopting the background models constructed from the simulations with monopole approximation, with the filled marks with dotted lines, i.e., squares for S12~GRm and diamonds for S20~GRm. As shown in \citetalias{SMT24}, the universal relation given by Eq.~(\ref{eq:STT21}) works well for the proto-neutron star oscillation frequencies determined with the Cowling approximation independently of the treatment of gravity whether the effective GR or the relativistic framework in the numerical simulations and also independently of the interpolation in the simulations. In addition to this universality, from this study, we find that the proto-neutron star oscillation frequencies determined with the Cowling approximation are well expressed with this relation even on the background models from the simulations with nonmonopole potential, although the frequencies given by this universal relation sometimes deviate from the gravitational wave signals in the simulations, depending on the set up of the simulations.

\begin{figure*}[tbp]
\begin{center}
\includegraphics[scale=0.6]{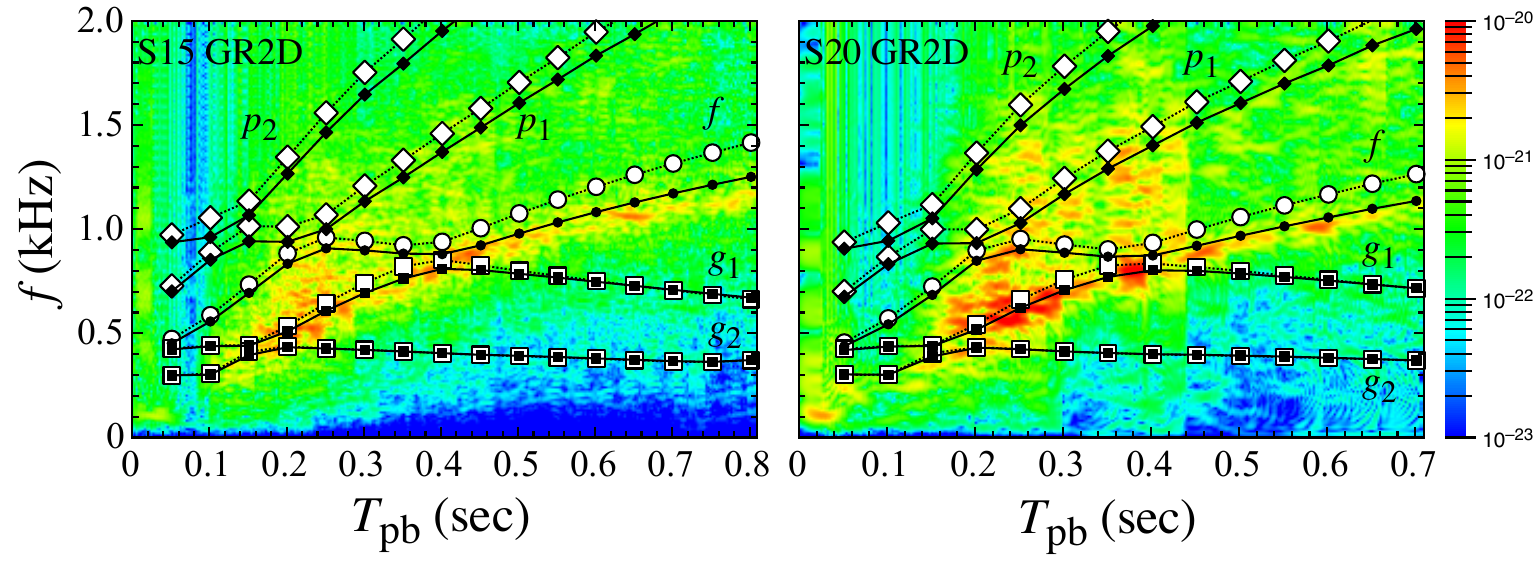}
\end{center}
\caption{
Same as Fig.~\ref{fig:GWSp_20SFHO_cow}, but also including the frequencies obtained with the metric perturbations (without the Cowling approximation) using the filled symbols. 
}
\label{fig:GWSp_20SFHO}
\end{figure*}

\begin{figure}[tbp]
\begin{center}
\includegraphics[scale=0.6]{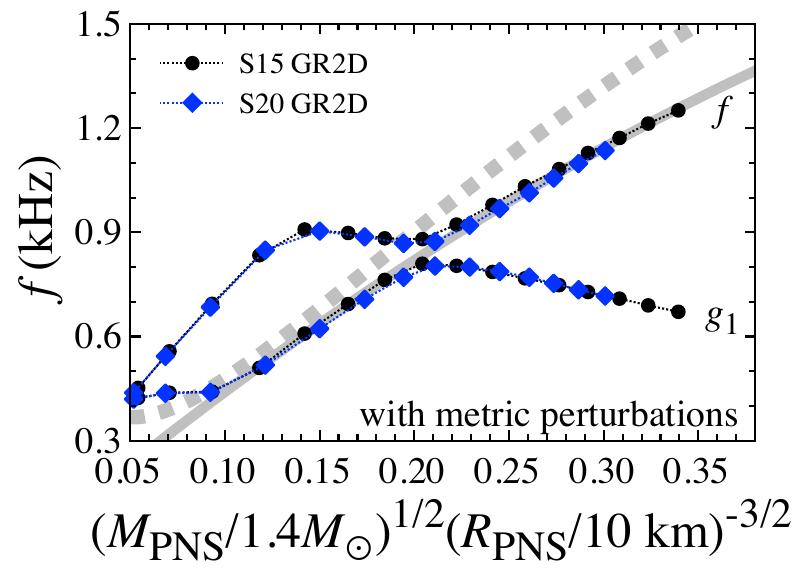}
\end{center}
\caption{
Same as Fig.~\ref{fig:fx-uni}, but also including the frequencies obtained without the Cowling approximation with the filled marks. The dotted line denotes the empirical relation given by Eq.~(\ref{eq:STT21}), while the solid line denotes the empirical relation given by Eq.~(\ref{eq:new_emp}).
}
\label{fig:fx-universal-new}
\end{figure}

Next, to try to recover the discrepancy between the gravitational wave signals in the simulations with non-monopole potential and the proto-neutron star oscillation frequencies, we examine the proto-neutron star oscillation frequencies without the Cowling approximation, i.e., including the metric perturbations, adopting the zero-damping approximation. In Fig.~\ref{fig:GWSp_20SFHO}, we show the results without the Cowling approximation with the filled marks with the solid lines, where the background contour denotes the gravitational wave signals appearing in the simulations, which are the same as the background contour shown in Fig.~\ref{fig:GWSp_20SFHO_cow}. From this figure, it is obvious that the gravitational wave signals in the simulations with non-monopole potential agree well with the proto-neutron star oscillation frequencies determined without the Cowling approximation. Furthermore, we find that the $f$- and $p_i$-modes without the Cowling approximation significantly deviate from the results with the Cowling approximation, while the deviation of the $g_i$-modes seems to be relatively small. Considering that the deviation exists even in the $g_i$-modes with the effective GR~\cite{ST2020c}, whether or not the deviation in the $g_i$-modes exists may also depend on the treatment of the gravity. Anyway, since the proto-neutron star oscillation frequencies without the Cowling approximation deviate from those with the Cowling approximation, we have to check the universality as a function of the proto-neutron star average density. 

Several studies have attempted to compare the gravitational wave spectra in the simulation and frequencies of the proto-neutron star oscillations determined with linear analysis, e.g., Fig.~5 in Morozova~{\it et al.}~(2018)~\cite{MRBV2018}, Fig.~11 with the Cowling approximation and Fig.~12 with non-Cowling in Torres-Forn\'{e}~{\it et al.}~(2019)~\citep{TCPOF2019a}, and Fig.~5 with the Cowling approximation in Sotani~and~Takiwaki~(2020) ~\cite{ST2020b} and Fig.~6 with non-Cowling in \cite{ST2020c}. However, it is not straightforward to interpret the results across these works. When compared to the oscillation frequencies under the Cowling approximation, the gravitational wave frequency is comparable in \cite{TCPOF2019a}, while higher in \cite{MRBV2018} and \cite{ST2020b}.
When compared to the non-Cowling model, the gravitational wave spectra are roughly comparable in \cite{TCPOF2019a,MRBV2018} but appear higher in \cite{ST2020c}. All of these studies use effective GR in their simulations (although the comparison with the general relativistic simulation is also shown in~\cite{TCPOF2019a}), while linear analysis is conducted in a general relativistic framework. 
As discussed in the Appendix in Ref.~\cite{ST2020b}, consistently constructing a background model from the effective GR simulation for the relativistic linear analysis remains difficult. This difficulty motivates our use of general relativistic simulations to generate consistent background models for analysis. 
Our results are along the same line with \citetalias{SMT24} and Zha {\it et al.} (2024)~\cite{Zha2024}.
Reinforcing the idea that employing consistent gravitational treatments in both simulations and linear analysis is key to minimizing discrepancies.

\begin{figure*}[tbp]
\begin{center}
\includegraphics[scale=0.6]{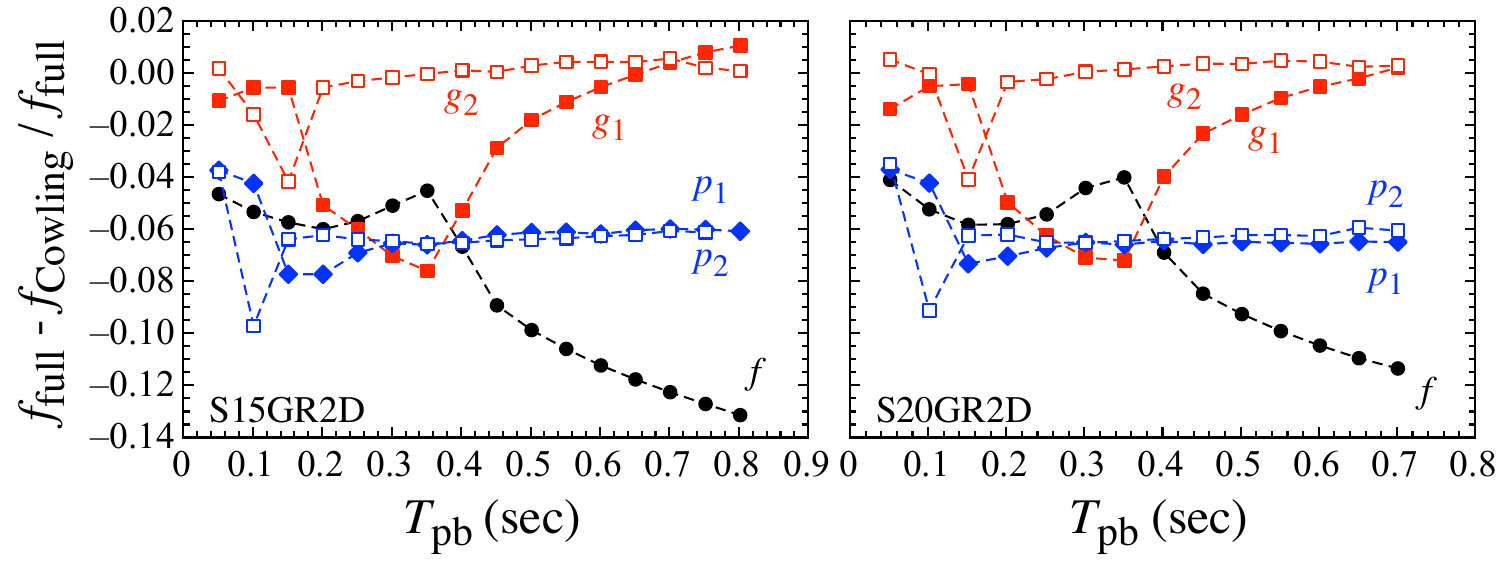}
\end{center}
\caption{
Time evolution of the relative deviation of the gravitational wave frequencies determined with the Cowling approximation from those with the metric perturbations. The left and right panels correspond to the results for the supernova model with S15GR2D and S20GR2D, respectively.
}
\label{fig:relative_devi}
\end{figure*}

\begin{figure}[tbp]
\begin{center}
\includegraphics[scale=0.5]{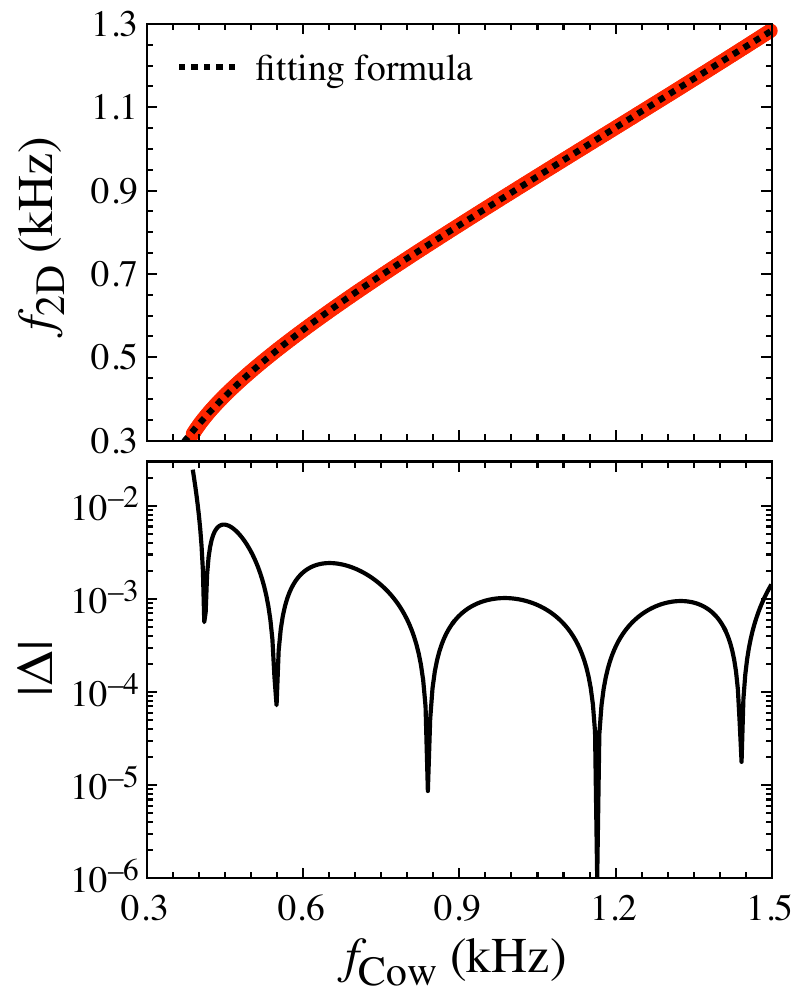}
\end{center}
\caption{
The gravitational wave frequencies with non-monopole gravity, $f_{\rm 2D}$, are shown in the upper panel as a function of the gravitational wave frequencies with the monopole approximation, $f_{\rm Cow}$, which corresponds to $x=0.07-0.35$. The dotted line denotes the fitting formula given by Eq.~(\ref{eq:fit_f2D}). In the bottom panel, the relative deviation from this fitting, defined by Eq.~(\ref{eq:Delta_f2D}), is shown.
}
\label{fig:f2D_fCow}
\end{figure}

In Fig.~\ref{fig:GWSp_20SFHO}, we plot the $f$- and $g_1$-mode frequencies determined with the metric perturbations as a function of the proto-neutron star average density, where the circles and diamonds denote the results with S15~GR2D and S20~GR2D. The thick dotted line is the frequencies calculated with the universal relation obtained with the Cowling approximation given by Eq.~(\ref{eq:STT21}), while the thick solid line is the fitting using the data of the frequencies with metric perturbation given as
\begin{equation}
   f_{\rm 2D} {\rm (kHz)} = 0.0082 + 4.5908x -2.6821x^2, \label{eq:new_emp}
\end{equation}
where $x$ is given by Eq.~(\ref{eq:xx}). The frequencies of the proto-neutron stars with metric perturbations significantly deviate from the universal relation derived with the Cowling approximation, while we find that the $f$- and $g_1$-mode frequencies corresponding to the gravitational wave signals in the simulation can be well fitted with the new fitting formula given by Eq.~(\ref{eq:new_emp})\footnote{
This new fitting formula is derived from the only two supernova models with different progenitor masses, using the same EOS. 
So, to verify the universality of this new relation, we have to do more simulations with various progenitor masses and EOS. In particular, since the explodability is not monotonic with the progenitor masses \cite{Muller16}, a careful study would be necessary.}.

The resultant new universal relation is illustrated in Fig.~\ref{fig:fx-universal-new}, where the $f$- and $g_1$-mode frequencies determined via the eigenvalue problem with the metric perturbations are compared with the empirical relation.
The thick gray solid line represents the new empirical relation derived from simulations that include metric perturbations~[Eq.~\eqref{eq:new_emp}], while the thick gray dashed line shows the previous relation based on the Cowling approximation~[Eq.~\eqref{eq:STT21}].

These curves provide a practical guideline for correcting Cowling-based frequency estimates to more accurate values obtained from fully relativistic perturbation analysis. In practice, Eq.~\eqref{eq:new_emp} can be directly applied using the proto-neutron star's mass and radius extracted from simulation profiles with non-monopole gravitational potential, offering a refined alternative to Eq.~\eqref{eq:STT21}. 
Notably, the Cowling-based relation in Eq.~\eqref{eq:STT21} was also obtained using simulations that employed phenomenological gravitational potential, as shown in the left panel of Fig.~10 in \citetalias{SMT24} (and also in Fig.~\ref{fig:fx-uni}).

In Fig.~\ref{fig:relative_devi}, we show the time evolution of the relative deviation of the gravitational wave frequencies determined with the Cowling approximation, $f_{\rm Cowling}$, from those with the metric perturbations, $f_{\rm full}$, for the supernova models S15GR2D (left panel) and S20GR2D (right panel). From this figure, it is found that the deviation in the $f$-mode frequency increases with time, while the deviation in the $g_2$-, $p_1$-, and $p_2$-modes frequencies are almost independent of post-bounce time. 

Using the fitting formulae derived from the gravitational wave frequecies with the Cowling approximation, $f_{\rm Cow}(x)$, given by Eq.~(\ref{eq:STT21}) and those with metric perturbations, $f_{\rm 2D}(x)$, given by Eq.~(\ref{eq:new_emp}), one can directly discuss the relation between $f_{\rm Cow}$ and $f_{\rm 2D}$ through the mediator variable, $x$, given by Eq.~(\ref{eq:xx}). That is, the gravitational wave spectra obtained from the numerical simulations with non-monopole gravity (corresponding to the frequencies with the metric perturbations) are estimated from those with monopole approximation (corresponding to the frequencies with the Cowling approximation) by using this relation. In practice, as shown in the upper panel of Fig.~\ref{fig:f2D_fCow}, one can derive the relation between $f_{\rm Cow}$ and $f_{\rm 2D}$, which can be fitted as a function of $f_{\rm Cow}$ given by 
\begin{align}
  f_{\rm 2D}^{\rm fit} =& 1.7800 + 0.9676 \ln(f_{\rm Cow}) -1.8052 f_{\rm Cow} \nonumber\\
  &+ 1.1441 f_{\rm Cow}^2 -0.2236 f_{\rm Cow}^3, \label{eq:fit_f2D}
\end{align}
as shown in the dotted line. The relative deviation defined as
\begin{equation}
  \Delta = \frac{|f_{\rm 2D}^{\rm fit} - f_{\rm 2D}|}{f_{\rm 2D}} \label{eq:Delta_f2D}
\end{equation}
is also shown in the bottom panel of Fig.~\ref{fig:f2D_fCow}, i.e., one can estimate $f_{\rm 2D}$ less than $\sim 1\%$ accuracy.



\begin{figure}[htbp]
\includegraphics[scale=0.43]{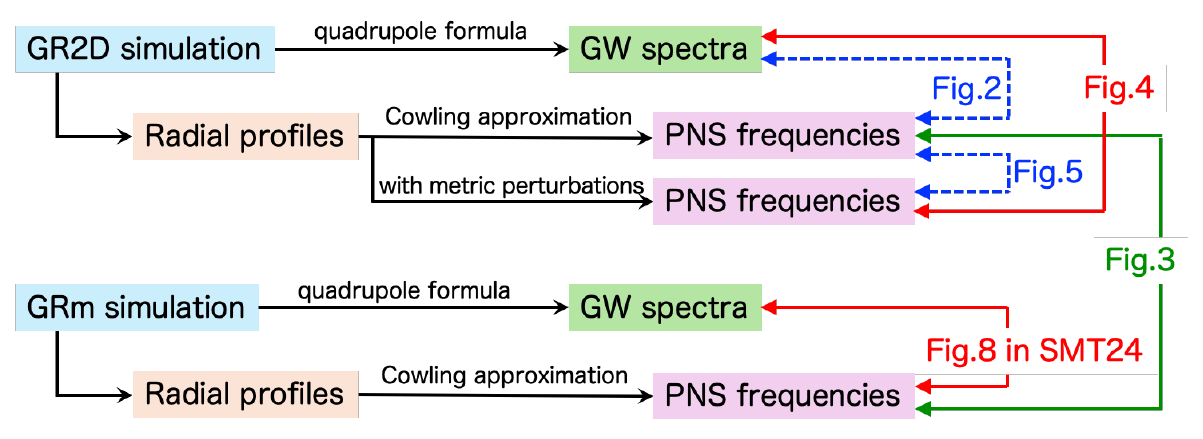}
\caption{
Summary of the treatment of metric perturbations in this work and our previous work.
The oscillation frequency of proto-neutron star (PNS) with metric perturbations with GR2D agrees with gravitational wave (GW) frequency in GR2D simulations (see Fig.~\ref{fig:GWSp_20SFHO}). 
The PNS frequency with Cowling approximation with GR2D overestimates the GW frequency in GR2D simulations (see Figs.~\ref{fig:GWSp_20SFHO_cow} and \ref{fig:fx-universal-new}). The solid and dashed lines denote that the results are consistent and inconsistent with each other, respectively.}
\label{fig:proceadure}
\end{figure}

At the end, we graphically summarize our methodology and key findings in Fig.~\ref{fig:proceadure}. Specifically, we examine five distinct approaches for estimating the gravitational wave frequency from the two classes of simulations: those employing a monopole gravitational potential (denoted GRm) and those incorporating a two-dimensional (non-monopole) gravitational treatment (denoted GR2D).
In each case, gravitational wave spectra are extracted using the quadrupole formula. Independently, we perform asteroseismological analyses on the proto-neutron star by solving for its eigenfrequencies, utilizing the radial profiles from the simulations. These oscillation frequencies are determined either under the Cowling approximation or by including full metric perturbations.
The most significant finding is that, in GR2D models, the gravitational wave frequencies obtained via the quadrupole formula are in excellent agreement with the proto-neutron star oscillation frequencies derived with full metric perturbations (see Fig.~\ref{fig:GWSp_20SFHO}). In contrast, the Cowling approximation systematically overestimates the gravitational wave frequencies in GR2D models, as shown in Figs.~\ref{fig:GWSp_20SFHO_cow} and \ref{fig:fx-universal-new}. In GRm models, the gravitational wave frequencies via the quadrupole formula and the proto-neutron star oscillaiton frequencies with the Cowling approximation are roughly consistent (see Fig.~8 in \citetalias{SMT24}).
Interestingly, the universal relation with the Cowling approximation, using the GRm results, is still held even for the proto-neutron star frequencies with the Cowling approximation, using the GR2D results (see Fig.~\ref{fig:fx-uni}). There is the caveat that the Cowling approximation in GR2D cannot capture the true gravitational wave frequency obtained by the quadrupole formula.
The agreement of the empirical relation under the Cowling approximation suggests that the radial profiles of GRm and GR2D models are similar when the proto-neutron star's mass and radius are fixed. To obtain frequencies that reflect full metric perturbations, we recommend using Eq.~\eqref{eq:new_emp}, with the proto-neutron star's mass and radius directly extracted from the simulation data.

\section{Conclusion}
\label{sec:Conclusion}

Core-collapse supernovae are a promising gravitational wave source next to the compact binary mergers. In the previous study, we have shown that the gravitational wave signals appearing in the numerical simulations with monopole approximation in the gravity agree with the $f$- (and $g_1$-) mode frequencies of proto-neutron stars determined with the Cowling approximation~\citepalias{SMT24}. However, in a realistic situation, the gravitational wave signals should be discussed with a numerical simulation with a nonmonopole gravitational potential. To discuss the supernova gravitational waves obtained in such a situation, we have newly performed the two-dimensional relativistic numerical simulations with non-monopole (two-dimensional) gravitational potential, adopting the $15M_\odot$ and $20M_\odot$ progenitor models with the SFHo EOS for the dense matter.

Then, we compare the gravitational wave signals in the simulations with the frequencies of proto-neutron stars determined with and without the Cowling approximation. As a result, we find that the frequencies of the proto-neutron star oscillations with the Cowling approximation overestimate the gravitational wave frequencies compared to the signals in the simulations, while those frequencies with the Cowling approximation are still well expressed in the universal relations obtained with the Cowling approximation, which is independent of the treatment of gravity whether the effective GR or relativistic framework in the simulations, the numerical interpolation, and whether the monopole or non-monopole gravitational potential. On the other hand, we find that the gravitational wave signals in the simulations with non-monopole potential agree well with the frequencies of the proto-neutron stars with the metric perturbations, while the resultant frequencies significantly deviate from the universal relation obtained with the Cowling approximation. Using the data we obtained in this study, we also derive the fitting formula expressing the gravitational wave frequencies in the simulations with a non-monopole potential, which is at least independent of the progenitor mass considered in this study. However, to verify the universality of this fitting formula, we have to do additional analysis with different supernova parameters, such as the progenitor mass and EOS of the dense matter, which will be done somewhere in the future. The key uncertainty is whether the frequency relation still works for very compact proto-neutron stars, beyond the mass range probed by the current models. Thus, testing the new universal relation for more massive progenitors 
($>20M_\odot$) with bigger cores that eventually approach or undergo black hole formation will be particularly important in the future.

Looking ahead, our findings not only reinforce the importance of sophisticated numerical methods in supernova simulations but also underscore the critical role of including metric perturbations in asteroseismological analyses.
Importantly, the revised universal relation proposed here offers a more reliable tool for probing the internal structure of proto-neutron stars. 
Considering the sensitivities of the gravitational-wave detectors~\cite{LIGO,VIRGO,KAGRA} improve in the future~\cite{ET,CE,NEMO,LRR20}, which enhances the likelihood of detecting Galactic supernovae, we anticipate that these theoretical advancements will complement the development of sophisticated data analysis pipelines~\cite{Bizouard21,Bruel23,Mukherjee2021,Hu2022,Sasaoka2023,Sasaoka2024,Powell2022,ACL23,Powell2024b}. Such efforts will be essential for extracting physical information, such as the equation of state, mass, and radius of proto-neutron stars, from observed signals. Ultimately, bridging detailed simulations and real observational data will deepen our understanding of stellar explosions and the fundamental physics governing dense matter.

\acknowledgments

This work is partly supported in part by Japan Society for the Promotion of Science (JSPS) KAKENHI Grant Number 
(JP22H01223,  
JP23K03400,  
JP23K20848, 
and JP24KF0090). 
This research is also supported by
MEXT as “Program for Promoting researches on the Supercomputer Fugaku” (Structure and Evolution of the Universe Unraveled by Fusion of Simulation and AI; Grant Number JPMXP1020230406) and JICFuS.
Part of the numerical computations was carried out on a PC cluster at Center for Computational Astrophysics, National Astronomical Observatory of Japan. BM acknowledges supported by the ARC through Discovery Project DP240101786. We acknowledge computer time allocations from Astronomy Australia Limited's ASTAC scheme, the National Computational Merit Allocation Scheme (NCMAS), and from an Australasian Leadership Computing Grant. Some of this work was performed on the Gadi supercomputer with the assistance of resources and services from the National Computational Infrastructure (NCI), which is supported by the Australian Government, and through support by an Australasian Leadership Computing Grant. 



\end{document}